\documentclass[12pt]{wlscirep}

\usepackage{setspace}
\title{Full-Field Interferometric Imaging of Propagating Action Potentials}

\author[1,2,*]{Tong Ling}
\author[3]{Kevin C. Boyle}
\author[1]{Georges Goetz}
\author[4]{Peng Zhou}
\author[2]{Yi Quan}
\author[5]{Felix S. Alfonso}
\author[3]{Tiffany W. Huang}
\author[1,2,*]{Daniel Palanker}
\affil[1]{Hansen Experimental Physics Laboratory, Stanford University, Stanford, CA 94305}
\affil[2]{Department of Ophthalmology, Stanford University, Stanford, CA 94305}
\affil[3]{Department of Electrical Engineering, Stanford University, Stanford, CA 94305}
\affil[4]{Department of Molecular and Cellular Physiology, Howard Hughes Medical Institute, Stanford University, Stanford, CA 94305}
\affil[5]{Department of Chemistry, Stanford University, Stanford, CA 94305}
\affil[*]{To whom correspondence may be addressed. Email: tongling@stanford.edu, palanker@stanford.edu}

\keywords{quantitative phase imaging, interferometric microscopy, action potential, membrane electromotility, optophysiology}

\begin{abstract}
\end{abstract}
\begin{document}

\flushbottom
\maketitle

\thispagestyle{empty}

\section*{Abstract}

Currently, cellular action potentials are detected using either electrical recordings or exogenous fluorescent probes sensing calcium concentration or transmembrane voltage. Ca imaging has low temporal resolution, while voltage indicators are vulnerable to phototoxicity, photobleaching and heating. Here we report full-field interferometric imaging of individual action potentials by detecting the movement across the entire cell membrane. Using spike-triggered averaging of the movies synchronized to electrical recording, we demonstrate deformations of up to 3 nm (0.9 mrad) during the action potential in spiking HEK-293 cells, with a rise time of 4 ms. The time course of the optically-recorded spikes matches electrical waveforms. Since the shot noise limit of the camera ($\sim$2 mrad/pix) precludes detection of the action potential in a single frame, for all-optical spike detection, images are acquired at 50 kHz, and 50 frames are binned into 1 ms steps to achieve a sensitivity of 0.3 mrad in a single pixel. Using self-reinforcing sensitivity enhancement algorithm based on iteratively expanding the region of interest for spatial averaging, individual spikes can be detected by matching the previously extracted template of the action potential with the optical recording. This allows all-optical full-field imaging of the propagating action potentials without exogeneous labels or electrodes.

\section*{Introduction}

Modern methods for detecting electrical activity in cells rely on either electrical or optical recordings, both of which are invasive. Electrical methods require electrodes being placed adjacent to the cells of interest \cite{Hamill1981, Margrie2002, Thomas1972, Csicsvari2003, Jones1992}. Optical measurements rely on exogenous fluorescent probes, such as calcium indicator \cite{Ohki2005} or trans-membrane voltage sensors \cite{Salzberg1983, Kralj2012, Hochbaum2014}. Ca imaging provides rather low temporal resolution \cite{Wilt2013}, while fluorescent voltage indicators can be phototoxic, and are limited by photobleaching \cite{Scanziani2009} and heating of the target \cite{Hochbaum2014}.

The large changes in trans-membrane voltage taking place during action potential have long been hypothesized to cause changes in the shape of biological cells, which is primarily determined by the balance of intracellular hydrostatic pressure, membrane tension, and strain exerted by the cytoskeleton \cite{Keren2008,Gauthier2012,Pierre2015}. A layer of mobile ions along the cell membrane exerts additional tension on the lipid bilayer due to their lateral repulsion \cite{Zhang2001,Holthuis2014,Savtchenko2017}, which makes membrane tension voltage-dependent (see Supplementary Fig. S1). A 100 mV depolarization during an action potential increases tension by about 10 $\mu$N/m \cite{Zhang2001}, which increases the force exerted on a membrane of a 10 $\mu$m cell by 0.3 nN. This is expected to deform the cell by decreasing its surface area while preserving the volume, thereby making it more spherical. The movement of the cell membrane, called electromotility \cite{Zhang2001, Petrov2002}, is expected to follow the voltage change nearly instantaneously, since the magnitude of this force is very significant at cellular scale: if it were not counteracted by the cytoskeleton, such a force would accelerate the cell by about 600 m/$s^2$---much more than is observed with 1 nm membrane displacement during 1 ms in action potential ($10^{-3}$ m/$s^2$).

Movements of the cell membrane that accompany action potentials have been detected in the giant squid axon and in crustaceans (0.3$\sim$5 nm) using the single-point reflection of a laser beam \cite{Hill1977,Fang-Yen2004,Akkin2009,LaPorta2012}, in a crab nerve (5$\sim$10 nm) by the shift of a light-obstructing target \cite{Iwasa1980}, and by atomic force microscopy \cite{Kim2007}. In mammalian cells, membrane electromotility in HEK-293 and PC-12 cells was measured with atomic force microscope and piezo sensors (1 nm displacement per 100 mV) \cite{Zhang2001, Nguyen2012}. The average displacement of the cell membrane was also detected using quantitative phase microscopy (QPM) \cite{Popescu2006, Bhaduri2014, Goetz2018} in HEK-293 cells who’s potential was periodically modulated by a voltage clamp \cite{Oh2012}. A recent publication about optical thickness fluctuations in neuronal cell culture measured with low coherence interferometry \cite{Batabyal2017} reports changes of an optical path difference (OPD) on the order of $\sim$2 nm. Since these measurements were performed in transmission, the actual cell membrane displacement is larger than OPD by the difference in refractive indices of the cell and the medium. With $n \sim 0.035$ \cite{Schurmann2016,Steelman2017,Locquin2013}, the membrane displacement should be $\sim$57 nm. This is nearly two orders of magnitude larger than the previous experimental results with neurons in culture \cite{Nguyen2012, Yang2018}. Besides, that study did not provide sufficient temporal resolution to assess whether these spikes are actually the action potentials, nor validated correlation of these fluctuations with action potentials using electrical recordings. Another recent publication describes membrane displacements ranging from 0.2 to 0.4 nm during action potential measured by averaging the changes in light intensity at the cell edge using bright-field microscope \cite{Yang2018}. However, possible fluid exchange between the cytoplasm and the patch clamp pipette may affect the extent of cellular movements in such experiments.

In this article, we demonstrate the dynamics of cellular movements during the action potentials propagating in cell culture, validated by electrical recordings, but without affecting the cellular processes. This technique, based on ultrafast Quantitative Phase Microscopy (QPM) with self-reinforcing sensitivity enhancement, enables label-free non-invasive optical approach to detection of individual action potentials. First, using simultaneous optical and electrical recordings by QPM and multielectrode array (MEA), we extract a spike-triggered average template of the optical phase changes in spiking HEK-293 cells during the action potentials. Individual spikes can then be detected just optically by matching this template with the phase images without the use of electrical recordings. The detection sensitivity is further enhanced by frame binning and by iterative spatial averaging of the expanding region of interest using a self-reinforcing lock-in algorithm.

\section*{Materials and Methods}

\subsection*{Simultaneous optical and electrical recording by QPM and MEA}

\begin{figure}[ht]
\centering
\includegraphics[width=\linewidth]{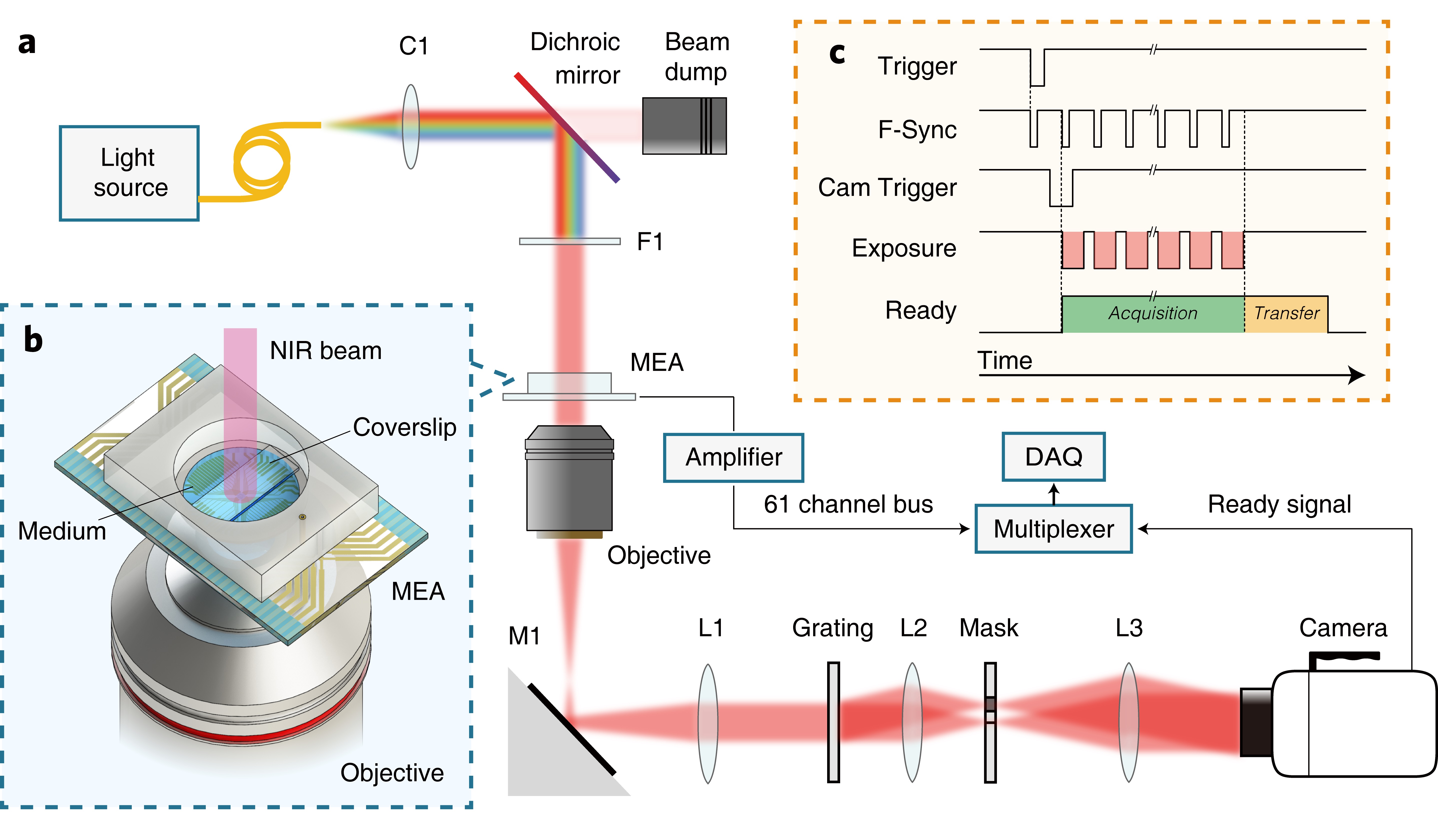}
\caption{System layout. {\bf(a)} Ultrafast QPM synchronized with MEA recording system. Light from a supercontinuum laser is collimated (C1) and filtered by a dichroic mirror and a band-pass filter (F1). An optical phase image of the sample is obtained from the off-axis interferogram captured by the high-speed camera. {\bf(b)} A transparent MEA plated with spiking HEK cells allows simultaneous near-infrared (NIR) optical recording and extracellular electrical recording. {\bf(c)} Electrical and optical measurements are synchronized by recording the camera “Ready” signal on one of the MEA channels. Trigger signal from the MEA and an external clock control the timing of the captured frames (see Methods).}
\label{fig:systemlayout}
\end{figure}

The setup for quantitative phase imaging was adapted from diffraction phase microscopy \cite{Bhaduri2014, Goetz2018} and is shown in Fig.~\ref{fig:systemlayout}. Recordings were initially performed at 1 kHz frame rate, using a fiber-coupled superluminescent diode (SLD, SLD830S-A20, Thorlabs, NJ) for illumination. For faster imaging (50 kHz), a supercontinuum laser (Fianium SC-400-4, NKT Photonics, Birkerød, Denmark) was used. In both cases, light from the fiber was collimated (SLD: F220FC-780, Thorlabs, NJ; supercontinuum laser had its built-in collimator) and the spectral components of interest were reflected towards the sample arm with a dichroic mirror (FF980-Di01-t1-25x36, Semrock, Rochester, NY). The wavelength range was further restricted to 797-841 nm by an optical bandpass filter (FF01-819/44-25, Semrock, Rochester, NY). Images formed by a 10$\times$ objective (CFI Plan Fluor 10X, N.A. 0.3, WD 16.0 mm, Nikon, Tokyo, Japan) and a 200 mm tube lens (Nikon, Tokyo, Japan) were projected onto a transmission grating (46-074, 110 grooves/mm, Edmund Optics, Barrington, NJ). The first diffraction order was passed through unobstructed, while the 0th order was filtered with a 150$\mu$m pinhole mask placed in the Fourier plane of a 4-f optical system, consisting of a 50 mm lens (AF Nikkor 50mm f/1.8D, Nikon, Tokyo, Japan) and a 250 mm bi-convex lens (LB1889-B, Thorlabs, NJ). Interferograms were formed on the camera sensor (Phantom v641, Vision Research, Wayne, NJ), which has a full well capacity of 11,000 electrons (digitized to 12 bit). At up to 1,000 fps, the camera can operate at a field size of up to 2560$\times$1600 pixels, while at 50,000 fps the field of view is reduced to 256$\times$128 pixels. To decrease the memory storage requirements, we used a field of view of 768$\times$480 pixels at 1,000 fps. External clock signal (Model 2100 Isolated Pulse Stimulator, A-M Systems, Sequim, WA) provided to the camera’s F-Sync input was triggered by the falling edge of a TTL trigger generated by the MEA. This trigger was also delivered to the camera via a digital delay generator (DG535, Stanford Research Systems, Sunnyvale, CA) to start image acquisition in synchrony with MEA recordings. The Ready output of the camera, high during both image acquisition (10 s for 768x$\times$480 pixels at 1,000 fps) and data transfer to the non-volatile memory of the camera ($\sim$8 s for 768$\times$480 pixels at 1,000 fps) was recorded by the data acquisition card (DAQ) of the MEA system to mark the start of each movie.

To retrieve the phase image, the Fourier transform of each interferogram was first calculated. The first diffraction order was then centered and isolated with a low-pass Gaussian filter. To monitor the changes in the phase image, the first interferogram in each movie sequence served as a reference. The phase difference between the reference and subsequent interferograms in the sequence was calculated by taking the complex angle of the pointwise division of the inverse Fourier transform of each filtered interferogram in the series and the filtered reference interferogram \cite{Pham2013}. Fluctuations resulting from $1/f$ noise of the illumination source were eliminated by subtracting the average of each phase image to achieve zero mean over the field of view (FOV). Highly noisy pixels, typically corresponding to a region obstructed by the electrodes of the MEA, were excluded from the analysis. The phase retrieval process was accelerated using a graphics processing unit (Tesla K40c, Nvidia, Santa Clara, CA).

Electrical signals were recorded using a custom 61-channel MEA system built on a transparent substrate with ITO leads \cite{Litke2004, Hottowy2012}. The recording electrodes were 10 $\mu$m in diameter and laid out in a hexagonal lattice with 30 $\mu$m spacing between neighboring electrodes and 30 $\mu$m spacing between rows. Platinum black was electro-deposited on the electrodes prior to every recording. The signals were amplified with a gain of 840 and filtered with a 43--2000 Hz bandpass. Signals were sampled at 20 kHz, using a National Instruments data acquisition card (NI PCI-6110, National Instruments, Austin, TX). The ready signal from the high-speed camera (Phantom v641, Vision Research, Wayne, NJ), marking the start of an image sequence acquisition, was used to synchronize electrical and optical recordings.

\subsection*{SNR optimization for spatial averaging}

Since some pixels display positive and others negative phase shift during action potential, proper spatial averaging should take into account polarity of the phase shift in each area:

\begin{equation}
\bar{\phi}(t) = \frac{1}{N} \sum_{ij}{\phi_{ij}(t) \cdot g_{ij}} ,
\end{equation}

\noindent where $N$ is the total number of pixels, $\phi_{ij}(t)$ is the phase shift at time $t$ in pixel ($i$,$j$) and $g_{ij}$ is the sign of the overall phase shift in that area. Note that this is different from averaging the absolute value of the phase, since $g_{ij}$ remain constant for each given pixel, while the sign of the noise changes over time. Averaging of the absolute values would not reduce the noise. A subset of the FOV can be selected to optimize the SNR of the spatially averaged phase signal based on the knowledge of maximum phase change during action potential in each pixel (signal amplitude $\Phi_{ij}$) and the noise level there ($\Delta_{ij}$). Both $\Phi_{ij}$ and $\Delta_{ij}$ are sorted across all pixels according to decreasing SNR into arrays $\Phi_k$ and $\Delta_k$. Then, the maximum SNR for spatial averaging can be calculated as following:

\begin{equation}
\overline{\textrm{SNR}}=\max_M{\frac{\sum_{k=1}^M{\Phi_k}}{\sqrt{\sum_{k=1}^M{\Delta_k^2}}}}, M \in [1,N] \cap \mathbb{Z^+} .
\end{equation}

\noindent The first $M$ pixels of the sorted arrays are then selected as the optimal subset for spatial averaging in that area.

\subsection*{Self-reinforcing lock-in spike detection}

\begin{figure}[ht]
\centering
\includegraphics[width=0.8\linewidth]{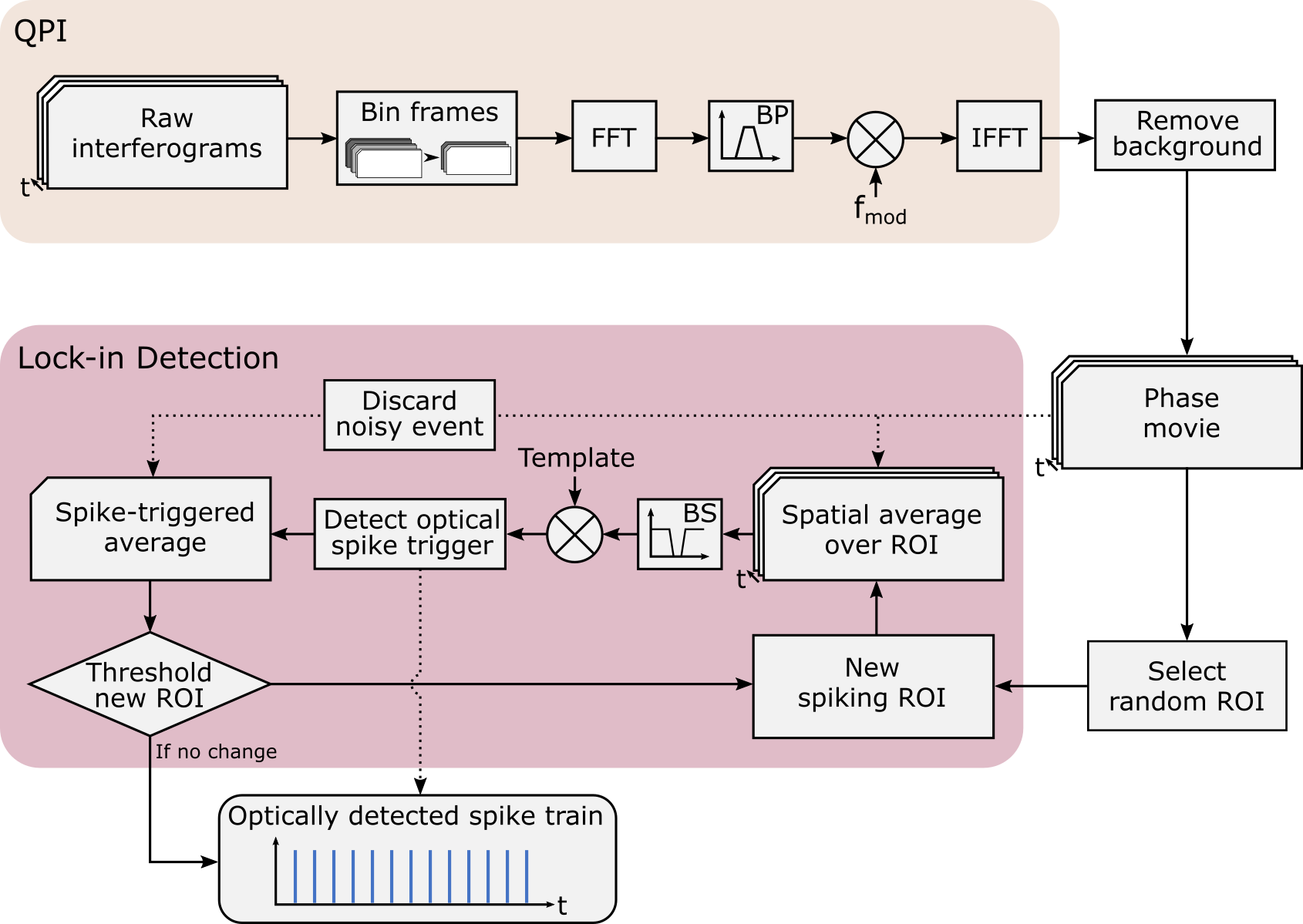}
\caption{Block diagram of QPM data processing and self-reinforcing lock-in algorithm for all-optical spike detection. After QPM processing, frame binning, and background removal, a random region of interest (ROI) is selected for the first iteration of the lock-in detection loop. The ROI is iteratively used to generate a spatial average of the spiking signal across the whole FOV, which is template-matched to generate an optical spike trigger. This trigger is used to form a STA from the original phase movie, and a new ROI is selected from the improved result. The process is repeated until the ROI stops changing between successive iterations.}
\label{fig:block}
\end{figure}

The SNR of individual pixels in the phase image is too low to reliably detect an action potential. However, since the spiking in a confluent culture of HEK cells is synchronized, spatial averaging can improve the SNR of the collective measurement. This spatial averaging must be applied with the positive and negative phase shifts across the cell taken into account, as described previously. However, since neither the distribution of the phase shift across the field of view, nor the spike timing in optical recordings are known \emph{a priori}, an iterative lock-in spike detection algorithm was developed to detect spikes in the noisy raw recording (Fig.~\ref{fig:block}).

Since the SNR of individual pixels is insufficient for reliably determining the sign of the phase shift at that location, initial spiking regions of interest (ROI) are chosen randomly. Phase changes are spatially averaged over the ROIs with a positive and negative sign randomly assigned to each pixel, yielding a single trace showing the displacement of the whole ROI. The randomly spatially averaged phase signal has a slightly improved SNR compared to single pixels. The resulting phase signal is filtered to remove mechanical vibrations and then cross-correlated with a characteristic template of the displacement during the action potential, which was obtained from a separate experiment using the reference MEA electrical recording to perform spatiotemporal spike-triggered averaging. The resulting cross-correlogram gives an estimate of the spikes timing in the spatially averaged phase signal, with spike times corresponding to peaks above a set prominence threshold. The detected spikes are used to create a spike-triggered average (STA) from the original movie, corresponding to a single action potential seen across the whole FOV. Each pixel in the STA movie is then correlated with the spike displacement template again to measure how similar that location’s displacement was to the template. The result provides an improved estimate of which parts of the FOV move together, and this ROI is used to start a new iteration of the loop. The process is repeated until the fraction of updated pixels between the new and old ROI decreases below a set threshold (200 pixels).

Each iteration of the lock-in detection algorithm improves the estimate of the ROI and thus the quality of the optically detected spike train and the STA movie. Fig.~\ref{fig:results}f shows that convergence happens in 4 iterations. The standard deviation of the delay between the detected optical and electrical spikes decreases with subsequent iterations (Fig.~\ref{fig:results}e), converging into a narrower distribution around zero delay (perfect detection). The timing of spikes detected in the initial iteration is nearly random, and contains large number of false positives and false negatives, but the final distributions are narrowly confined with an 11.6 ms standard deviation.

\subsection*{Sample preparation}

Spontaneously spiking HEK cells expressing Na\textsubscript{v} 1.3 ion channel were originally developed by Adam Cohen's group at Harvard University \cite{Park2014}. The cells were grown in a 1:1 mixture of Dulbecco's Modified Eagle Medium and F-12 supplement (DMEM/F12). The medium contained 10\% fetal bovine serum, 1\% penicillin (100 U/mL), streptomycin (100 $\mu$g/mL), geneticin (500 $\mu$g/mL), and puromycin (2 $\mu$g/mL). In order to spontaneously spike, HEK cells need to express not only Na\textsubscript{v} 1.3, but also the K\textsubscript{ir} 2.1 ion channel. Hence, they were transfected with the plasmid pIRES-hyg-K\textsubscript{ir}2.1 AMP Resistance using CalFectin as the transfection reagent. 30 minutes after a medium change, 3$\mu$L of 2.2 $\mu$g/$\mu$L of the plasmid was mixed with 1 mL DMEM. Then, 3 $\mu$L of CalFectin was added to the solution and 10-15 minutes later, the mixture was added to the cell culture. About 6 hours later, the cell culture was replaced, and the cells were used in experiments 24 hours later.

Spiking HEK-293 cells were plated on the MEA coated with Poly-D-Lysine (P6407, Sigma-Aldrich) with a density of 2,000 cells / mm$^2$ one day before the recording. Culture medium (DMEM + 10\% FBS) filtered with a sterile vacuum filter (SCGP00525, EMD Millipore, Darmstadt, Germany) was used to wash away any floating particles in the MEA chamber, using 2 applications of 800 $\mu$L each time. Two hours prior to recording, 7/8\textsuperscript{th} of the culture medium was replaced with the recording medium (Tyrode's solution, in mM: 137 NaCl, 2.7 KCl, 1 MgCl\textsubscript{2}, 1.8 CaCl\textsubscript{2}, 0.2 Na\textsubscript{2}HPO\textsubscript{4}, 12 NaHCO\textsubscript{3}, 5.5 D-glucose). 1/8\textsuperscript{th} of the culture medium was kept to avoid osmotic shock. Extra recording medium was then aspirated until the fluid just filled the 1.5 mm gap between a coverslip and the MEA (see the diagram in Fig.~\ref{fig:systemlayout}b and the actual bright-field image of the confluent HEK cells on MEA in Supplementary Fig. S1). Temperature was maintained at 29 $^\circ$C during the recordings.

\subsection*{Whole-cell patch clamp}

Spiking HEK-293 cells were placed in the same bath solution as used for QPI at room temperature (25 $^\circ$C). Glass micropipettes were pulled from borosilicate glass capillary tubes (Warner Instruments) using a PC-10 pipette puller (Narishige) and were loaded with internal solution, containing (in mM) 125 potassium gluconate, 8 NaCl, 0.6 MgCl\textsubscript{2}, 0.1 CaCl\textsubscript{2}, 1 EGTA, 10 HEPES, 4 Mg-ATP, 0.4 Na-GTP (pH 7.3, adjusted with NaOH; 295 mOsm, adjusted with sucrose). The resistance of pipettes filled with internal solution varied between 2 and 3~M$\Omega$. After setting up the whole cell configuration, the membrane potential during spontaneous action potentials of spiking HEK cells was monitored with a Multiclamp 700B amplifier (Molecular Devices) under the current clamp mode.

\section*{Results}

\subsection*{Dynamics of cellular deformation during action potential}

\begin{figure}[ht!]
\centering
\includegraphics[width=\linewidth]{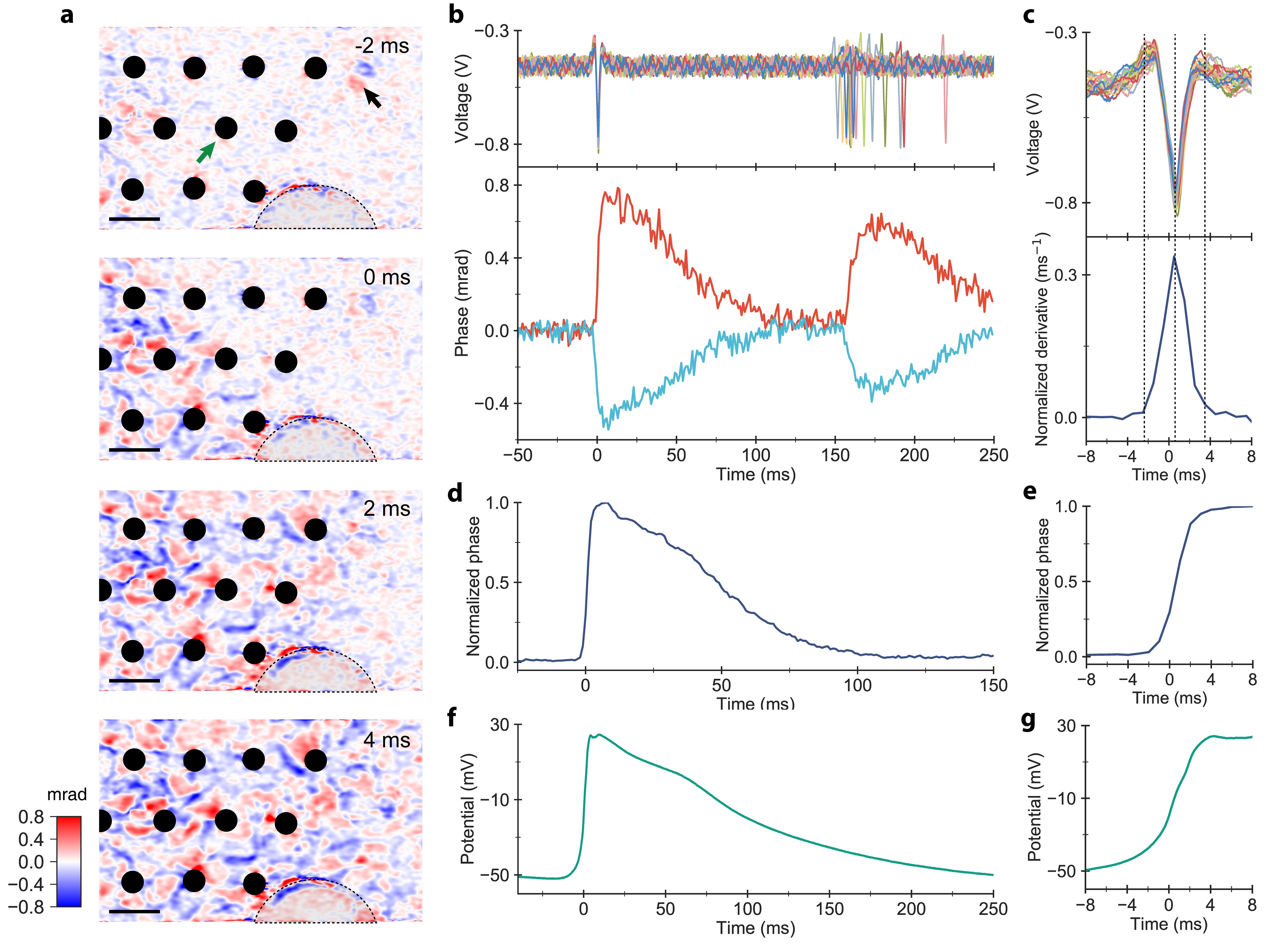} 
\caption{Spike-triggered average of the optical phase shift and electrical signal during action potential, obtained by averaging 5130 events. {\bf(a)} Propagation of the action potential across the field of view over 6 ms (see Supplementary Video 1). Timing of the frames is shown relative to the electrical spike detected on electrode indicated by a green arrow. Phase changes can be both positive and negative (shown in false color). The black arrow points at a floating cell, which produced larger phase shift than that of the action potential, while the dashed line outlines a semi-circular section with detached cells. Scale bar: 25 $\mu$m. {\bf(b)} Top row: Electrical signals recorded on the reference electrode, and aligned to the time of maximum deflection. Subsequent spikes exhibit some degree of natural jitter. Bottom row: Average optical phase signals extracted from two individual pixels near the reference electrode at the time of an electrical action potential. {\bf(c)} Comparison between the electrical signal on the reference electrode (top row) and the time derivative of the optical signal (bottom row). Normalized optical phase signal spatially averaged across the whole FOV {\bf(d)} and its rising edge {\bf(e)}, with spike timing corrected by local delays relative to the spike on the reference electrode. Patch clamp recording of the membrane potential {\bf(f)} and its rising edge {\bf(g)}.}
\label{fig:sta}
\end{figure}

To image cellular deformations during action potentials, we cultured genetically modified HEK-293 cells expressing a voltage-gated sodium channel (Na\textsubscript{V}1.3) and potassium channel (K\textsubscript{ir}2.1) \cite{Park2014, McNamara2016}. These cells spontaneously spike in synchrony, when cultured confluently \cite{Park2014}. For simultaneous electrical and optical recording, cells were plated on a transparent multielectrode array (MEA) at 90\% confluence (see Methods and Fig.~\ref{fig:systemlayout}b). The sample was illuminated by a superluminescent diode (SLD) with 1 mW/mm$^2$ irradiance. The camera frame rate was 1 kHz, with exposure duration of 45 $\mu$s, and field of view (FOV) on the sample was 159$\times$99 $\mu$m$^2$. The spontaneous spiking rate of the HEK cells was about 5 Hz, and recordings were conducted over 17 minutes in order to acquire about five thousand spikes. Phase images were retrieved from QPM interferograms using Fourier-domain processing (see Methods). The average optical phase movie recorded during the 50 ms preceding a spike, and up to 250 ms after it, which we call the spike-triggered average (STA) optical phase recording, was created by averaging movies from 5130 events (Fig.~\ref{fig:sta}, Supplementary Video 1) aligned in time on the basis of the electrical recordings of the action potentials by MEA. This movie demonstrates a rapid ($\sim$4 ms) rise of the phase in cells during the action potential, followed by a gradual ($\sim$100 ms) decline back to baseline. Extension of the averaging time window beyond the inter-spike duration, to include the following spike, demonstrates that natural jitter between the spontaneous spikes results in smearing of the average. As illustrated in Fig.~\ref{fig:sta}, some pixels exhibit positive change in phase, and some negative, corresponding to an increase and decrease in the cell thickness, respectively. The action potential wavefront propagated across the FOV in about 6 ms (Fig.~\ref{fig:sta}a), corresponding to 27 mm/s velocity. In some cells, optical phase increased on one side and decreased on the other, while in others, it increased at the center and decreased along the boundaries. The maximum amplitude of the positive phase shift was $\Delta\phi=0.86$ mrad, corresponding to a $\Delta h \sim$3.2 nm increase in cell thickness ($\Delta \phi = 2 \pi / \lambda \cdot \Delta n \Delta h$), assuming a refractive index difference of $\Delta n$ = 0.035 between the cytoplasm ($n\sim$1.37 \cite{Schurmann2016, Steelman2017}) and the cell culture medium ($n$ = 1.335 for Tyrode’s solution \cite{Locquin2013}). This value of $n$ also matches the 3.5 rad phase shift in the $\sim$13~$\mu$m thick spiking HEK-293 cell shown in the Supplementary Fig. S2.

By adjusting the timing of the video frames to the local timing of the action potential recorded on MEA, phase changes in all pixels can be averaged together for further improvement in SNR. The average optical waveform recorded during an action potential, shown in Fig.~\ref{fig:sta}d, has a SNR of 47.7 dB (243:1), and is very similar to the shape of the membrane potential recorded with a whole-cell patch clamp (Fig.~\ref{fig:sta}f,g). Note that the tail of the action potential in the patch clamp recording is a bit longer, likely due to differences in temperature and confluence of the cell culture in these preparations \cite{Park2014} (see Methods). General similarity of these waveforms indicates that cellular deformation during action potential reflects the changes in trans-membrane voltage. In extracellular recordings, the cells and electrodes are capacitively coupled, therefore the electrical signals correspond to the negative derivative of the intracellular voltage, bandpass filtered to the 43--2000 Hz range. As can be seen in Fig.~\ref{fig:sta}c, time derivative of the rising edge of the optically-recorded action potential (bottom frame) matches the timing and duration of the electrical spike (top frame), while derivative of the slow falling edge of the action potential was filtered out by the MEA.

\subsection*{Noise reduction and template matching for single-spike detection}

To reduce the noise for single-spike detection, global fluctuations in the phase image, which originate from mechanical vibrations of the optical components and $1/f$ noise of the light source, were removed by subtracting the mean value from every frame of the phase image (see Methods). This reduced variations to the shot noise level, set by the well capacity of the camera pixels \cite{Hosseini2016}, with the temporal standard deviation of the phase in a single pixel decreasing from $\sim$3.8 mrad to $\sim$1.9 mrad (see Supplementary Fig. S3). However, this value remains much larger than the maximum phase change during an action potential (0.9 mrad), and additional improvements are needed to achieve single-spike detection.

\begin{figure}
\centering
\includegraphics[width=\linewidth]{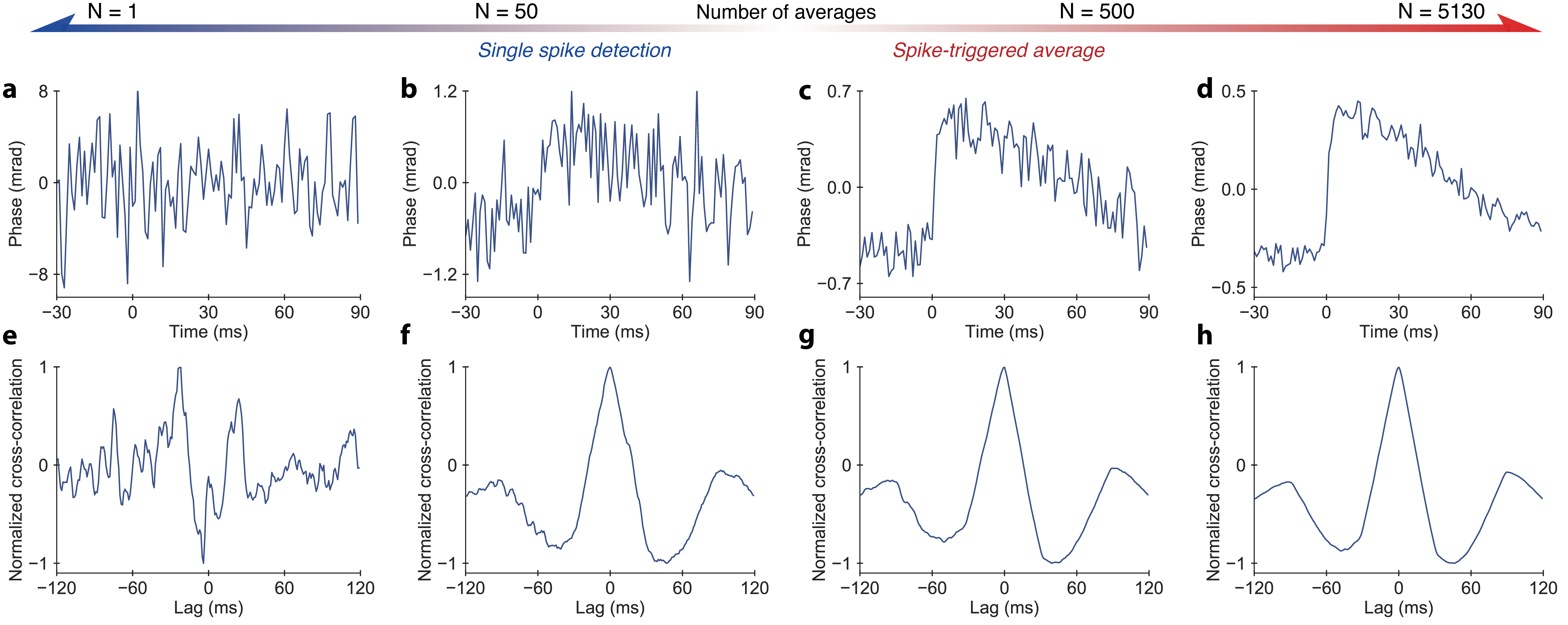}
\caption{Phase changes in a single pixel as a function of the number of averaged spikes, and corresponding cross-correlation with a phase template of the action potential. Top row ({\bf{a}}-{\bf{d}}): When the number of averaged spikes increases from $N=1$ ({\bf{a}}) to $N=5130$ ({\bf{d}}), the SNR of the phase change increases approximately as $\sqrt{N}$. Bottom row ({\bf{e}}-{\bf{h}}): Cross-correlation of the phase trace with the spike template shown in Fig. \ref{fig:sta}d illustrates that a spike can be detected from 50 averages, but not from a single trace. Additional averaging marginally increases the SNR of the cross-correlation. In the left two columns, averages can still be performed by frame binning using an ultrafast camera for single spike detection, while the right two columns illustrate the high fidelity detection of the cellular movement based on larger number of averages using STA.}
\label{fig:avgbinning}
\end{figure}

Using frame binning at higher imaging rate can increase the SNR of the recording, provided sufficient illumination intensity for shorter exposures. In the shot-noise limited regime, when readout noise is negligible, binning $N$ frames into one reduces the phase noise by approximately $\sqrt{N}$. With $N=50$ frames averaged, the noise in a single pixel decreased to $\sim$0.3 mrad, as shown in Supplementary Fig. S3 and Fig.~\ref{fig:avgbinning}. Further improvement in spike detection was achieved using template matching.  When implemented as a matched filter, the output SNR can be increased to $E/2N_0$, where $E$ is the total signal energy of the template and $N_0$ is the power spectral density (W/Hz) of the noise in the original signal before the matched filtering \cite{Richards2005}. The effect of temporal averaging by summation of separate spikes, synchronized via electrical recordings, is illustrated in Fig.~\ref{fig:avgbinning}a-d. Fig.~\ref{fig:avgbinning}b and f shows how an average of $N=50$ separate spikes, matched to the template shown in Fig.~\ref{fig:sta}d, clearly identifies a spike with no lag (0 ms). In contrast, the template match with no temporal averaging is insufficient (Fig.~\ref{fig:avgbinning}e). Averaging of larger numbers of spikes, shown in Fig.~\ref{fig:avgbinning}c-d, helps to further increase the SNR, but it does not significantly improve the template matching precision (Fig.~\ref{fig:avgbinning}g-h).

\begin{figure}[ht!]
\centering
\includegraphics[width=0.8\linewidth]{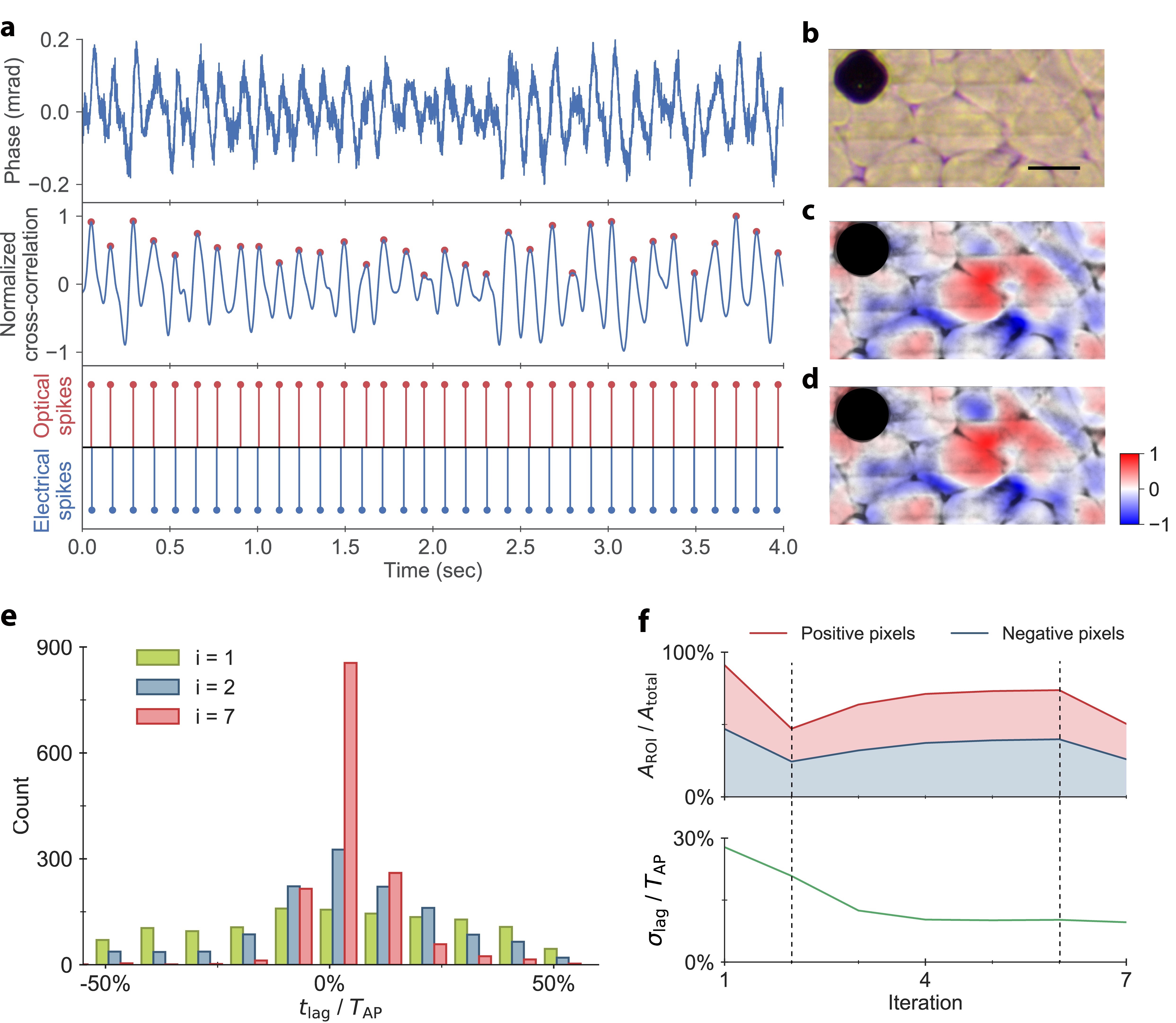}
\caption{All-optical detection of single action potential by interferometric imaging using a self-reinforcing lock-in algorithm. {\bf(a)} Spatially averaged phase signals, in the final iteration of the lock-in algorithm, showing a periodic phase signal (top row). Template matching is implemented by cross-correlating this phase signal with an action potential template obtained in a separate recording, and peaks are identified as spikes (middle row). Timing of the optically and electrically detected spikes is compared in the bottom row. 1584 spikes were recorded electrically in this FOV.  {\bf(b)} Spiking HEK-293 cells grown on the MEA, as seen in a bright-field microscope. The {\bf(c)} electrically- and {\bf(d)} optically-synchronized spike-triggered average movies cross-correlated with the action potential phase template demonstrate nearly identical spike correlation maps during the action potential. {\bf(e)} Histograms of the time lag between the optically and electrically detected spikes after 1, 2 and 7 iterations show convergence to a distribution centered around zero (mean 3.6\%, and standard deviation 9.7\% of $T_{AP}$). Analysis is based on 1584 spikes recorded electrically in this FOV. {\bf(f)} Convergence analysis. The dashed lines indicate successive three stages of the algorithm. During the 1$^{\textrm{st}}$ iteration, a ROI is selected randomly. In the second stage, from the 2$^{\textrm{nd}}$ iteration to steady-state (6$^{\textrm{th}}$ iteration here) the ROIs are refined and extended to finally cover $>$70\% of the FOV. In the final iteration (7$^{\textrm{th}}$ iteration here), the ROI is further re-charted based on SNR optimization for spatial averaging. The area chart (top row) shows the evolution of the ratio of ROI area to the total area. The bottom row shows the standard deviation of the time lags between the optically and electrically detected spikes, which converges to 9.7\% of the action potential period ($T_{AP} \sim$120 ms).}
\label{fig:results}
\end{figure}

\subsection*{All-optical spike detection by self-reinforcing sensitivity enhancement}

To demonstrate all-optical spike detection with QPM, we recorded spontaneously spiking HEK-293 cells at 50 kfps within a smaller FOV of 53$\times$26.5$\mu$m$^2$, and obtained ground truth electrical spiking data from the MEA for comparison. To achieve sufficiently short exposures, we used a supercontinuum laser providing irradiance of 4.7 mW/mm$^2$ (797--841 nm wavelength) on the sample, sufficiently bright for 10 $\mu$s exposures, thus enabling a camera frame rate up to 100 kHz. For all-optical spike detection, i.e. imaging of the action potentials without electrical recording, we developed an iterative lock-in algorithm that matches phase shift signals to a spike template that was previously recorded in a different set of HEK cells with the aid of MEA (Fig.~\ref{fig:sta}d). Using only this template, the algorithm iteratively estimates (a) the timing of action potentials for temporal averaging and (b) regions of interest that spike with the same polarity for spatial averaging (see Methods). After four iterations, the spike timing and spatial distribution of the phase shift polarity stabilize, with electrically- and optically-triggered average STA maps becoming very similar (Fig.~\ref{fig:results}c-d and Supplementary Video 2). The time course of the phase changes averaged over the FOV, shown in the top row of Fig.~\ref{fig:results}a, exhibit SNR of $\sim$20 dB. After cross-correlation with the spike template (middle row in Fig.~\ref{fig:results}a), peaks corresponding to the spike timing can be clearly identified (red dots). Timing of these peaks closely matches the spike timing identified in electrical recording, as shown in the bottom row of Fig.~\ref{fig:results}a. The quality of the spike detection is summarized by the standard deviation of the time lags between the optically and electrically detected spikes (Fig.~\ref{fig:results}e). This distribution starts out flat, showing nearly random spike detection in the first iteration, and quickly narrows around zero delay (perfect detection) with a standard deviation of about 11.6 ms, corresponding to 9.7\% of the action potential period ($\sim$120 ms). The optical spikes counted in this distribution are only those that were uniquely associated with an electrical spike within one action potential period. Other, rarer occurrences, where two or more optical spikes appeared within one electrical period, were considered false positives, and occurred at a false discovery rate of 0.07\%. The rate of false negative events (absence of any optical spike during an electrical spike) was 8.5\%. In addition to all-optical spike detection in the whole FOV, single spikes of individual cells can also be obtained within the cell boundaries segmented by the brightfield microscope image.

\section*{Discussion}

Mechanisms behind the optical phase change during action potential have been actively debated in literature. One proposed explanation was a change in refractive index associated with ions flowing into the cell during the action potential, or cell swelling due to water molecules accompanying those ions \cite{Kim2007, Oh2012}. The number of Na\textsuperscript{+} ions entering the cell during the depolarization phase of an action potential is: $N=C_m A \Delta V_m /e$, where $\Delta V_m \sim$100 mV is the transmembrane potential rise, $C_m$ is the specific membrane capacitance ($\sim 0.5 \mu$F cm$^{-2}$ \cite{Zhang2001}), $A$ is the cell membrane surface area, and $e$ is the elementary charge. For a cell of 10 $\mu$m in diameter, N is about 106, which is less than $3\times 10^{-4}$ of the number of Na\textsuperscript{+} ions in a mammalian cell of this size. Since the change in refractive index is proportional to the variation of the ion concentration \cite{Berlind2008}, and Na\textsuperscript{+} represents less than 10\% of the total amount of ions in a cell, the associated phase change for a cell having optical thickness of about 3 rad (Supplementary Fig. S2) is not expected to exceed 0.1 mrad---about an order of magnitude below the changes we observed during action potential. Furthermore, since four water molecules comprise the hydration coat of a sodium ion \cite{Catterall2017}, a total of about $4\times 10^6$ water molecules enter the cell with Na\textsuperscript{+} ions during the action potential. They increase the cell volume by about a factor of $2.3\times 10^{-7}$, or its diameter by about $0.7\times 10^{-3}$ nm, which is 3 orders of magnitude less than we observed during the action potential. It follows that neither the change in refractive index due to the influx of ions, or cell swelling from water accompanying those ions, can account for the observed phase changes.

Another proposal was that cell swelling might be caused by water diffusion associated with osmotic changes during action potential. However, this process is relatively slow: mouse cortical neurons swell in a hypotonic solution (144 mOsm/kg H2O, compared to 229 mOsm/kg H\textsubscript{2}O in the standard perfusion medium) with a time constant of $\sim$30 s \cite{Rappaz2005}---much longer than the few ms rise time of the action potential. Moreover, dilution of the cellular content by influx of water would result in a decrease of the optical path at the center of a spherical cell, and an increase at the boundaries, due to cell expansion \cite{Rappaz2005, Boss2013}. Our observations were the opposite: optical phase typically increased at the center and decreased along the cell boundaries. Therefore, changes in the refractive index due to cell swelling from water diffusion is unlikely to be the mechanism behind the rapid phase changes we observed. Cellular deformation due to an increase in membrane tension during depolarization is a more likely explanation of the observed ms-scale dynamics, including the positive and negative phase changes within individual cells.

In summary, our observations demonstrate that mammalian cells deform during the action potential by about 3 nm, and that the dynamics of this deformation matches the time course of the changes in cell potential. Phase changes associated with an action potential, measured in transmission, are below the shot-noise floor in a single frame. However, with sufficient temporal and spatial averaging, and prior knowledge about the spike shape, leveraged through template matching in our study, cellular deformations during a single action potential can be detected. In principle, phase changes could be increased using a multi-pass cavity \cite{Juffmann2016}, although it would require a light source with sufficiently long coherence length, which is likely to increase speckling in the interferograms. SNR could also be improved by reducing the shot noise using brighter illumination in conjunction with a camera having higher sampling rate and/or larger well capacity \cite{Hosseini2016}. Even so, as implemented, this system achieves all-optical, label-free, full-field imaging of electrical activity in mammalian cells based on high-speed QPM, and may enable non-invasive optophysiological studies of neural networks.

\section*{Acknowledgements}
Funding was provided by the NIH grant U01 EY025501 and by the Stanford Neurosciences Institute (G.G.).

\noindent We would like to thank Dr. A. Roorda and Dr. B.H. Park for helpful comments and encouragement, Dr. E.J. Chichilnisky for providing the MEA, Thomas Flores for assistance with MEA setup, and Yijun Jiang for help with the data processing and theoretical analysis. We are also grateful to laboratory of Bianxiao Cui for their help with transfection and culturing of the spiking HEK cells.

\section*{Author contributions}
G.G. and D.P. designed the system. T.L. and G.G. conducted the experiments. T.L. and K.C.B. analyzed the data. Y.Q., F.S.A. and T.H. prepared cell cultures. T.L., K.C.B., G.G. and D.P. wrote the manuscript. All work was supervised by D.P.

\noindent \emph{Conflict of interest statement}: All authors declare no competing financial interests.

\bibliography{references}

\begin{thebibliography}{10}
\expandafter\ifx\csname url\endcsname\relax
  \def\url#1{\texttt{#1}}\fi
\expandafter\ifx\csname urlprefix\endcsname\relax\def\urlprefix{URL }\fi
\expandafter\ifx\csname doiprefix\endcsname\relax\def\doiprefix{DOI }\fi
\providecommand{\bibinfo}[2]{#2}
\providecommand{\eprint}[2][]{\url{#2}}

\bibitem{Hamill1981}
\bibinfo{author}{Hamill, O.~P.}, \bibinfo{author}{Marty, A.},
  \bibinfo{author}{Neher, E.}, \bibinfo{author}{Sakmann, B.} \&
  \bibinfo{author}{Sigworth, F.~J.}
\newblock \bibinfo{journal}{\bibinfo{title}{Improved patch-clamp techniques for
  high-resolution current recording from cells and cell-free membrane
  patches}}.
\newblock {\emph{\JournalTitle{Pfl\"ugers Archiv}}}
  \textbf{\bibinfo{volume}{391}}, \bibinfo{pages}{85--100}
  (\bibinfo{year}{1981}).

\bibitem{Margrie2002}
\bibinfo{author}{Margrie, T.~W.}, \bibinfo{author}{Brecht, M.} \&
  \bibinfo{author}{Sakmann, B.}
\newblock \bibinfo{journal}{\bibinfo{title}{In vivo, low-resistance, whole-cell
  recordings from neurons in the anaesthetized and awake mammalian brain}}.
\newblock {\emph{\JournalTitle{Pfl\"ugers Archiv}}}
  \textbf{\bibinfo{volume}{444}}, \bibinfo{pages}{491--498}
  (\bibinfo{year}{2002}).

\bibitem{Thomas1972}
\bibinfo{author}{Thomas, C.~A.}, \bibinfo{author}{Springer, P.~A.},
  \bibinfo{author}{Loeb, G.~E.}, \bibinfo{author}{Berwald-Netter, Y.} \&
  \bibinfo{author}{Okun, L.~M.}
\newblock \bibinfo{journal}{\bibinfo{title}{A miniature microelectrode array to
  monitor the bioelectric activity of cultured cells}}.
\newblock {\emph{\JournalTitle{Experimental Cell Research}}}
  \textbf{\bibinfo{volume}{74}}, \bibinfo{pages}{61--66}
  (\bibinfo{year}{1972}).

\bibitem{Csicsvari2003}
\bibinfo{author}{Csicsvari, J.} \emph{et~al.}
\newblock \bibinfo{journal}{\bibinfo{title}{Massively parallel recording of
  unit and local field potentials with silicon-based electrodes}}.
\newblock {\emph{\JournalTitle{Journal of Neurophysiology}}}
  \textbf{\bibinfo{volume}{90}}, \bibinfo{pages}{1314--1323}
  (\bibinfo{year}{2003}).

\bibitem{Jones1992}
\bibinfo{author}{Jones, K.~E.}, \bibinfo{author}{Campbell, P.~K.} \&
  \bibinfo{author}{Normann, R.~A.}
\newblock \bibinfo{journal}{\bibinfo{title}{A glass/silicon composite
  intracortical electrode array}}.
\newblock {\emph{\JournalTitle{Annals of Biomedical Engineering}}}
  \textbf{\bibinfo{volume}{20}}, \bibinfo{pages}{423--437}
  (\bibinfo{year}{1992}).

\bibitem{Ohki2005}
\bibinfo{author}{Ohki, K.}, \bibinfo{author}{Chung, S.},
  \bibinfo{author}{Ch'ng, Y.~H.}, \bibinfo{author}{Kara, P.} \&
  \bibinfo{author}{Reid, R.~C.}
\newblock \bibinfo{journal}{\bibinfo{title}{Functional imaging with cellular
  resolution reveals precise micro-architecture in visual cortex}}.
\newblock {\emph{\JournalTitle{Nature}}} \textbf{\bibinfo{volume}{433}},
  \bibinfo{pages}{597} (\bibinfo{year}{2005}).

\bibitem{Salzberg1983}
\bibinfo{author}{Salzberg, B.~M.}, \bibinfo{author}{Obaid, A.~L.},
  \bibinfo{author}{Senseman, D.~M.} \& \bibinfo{author}{Gainer, H.}
\newblock \bibinfo{journal}{\bibinfo{title}{Optical recording of action
  potentials from vertebrate nerve terminals using potentiometric probes
  provides evidence for sodium and calcium components}}.
\newblock {\emph{\JournalTitle{Nature}}} \textbf{\bibinfo{volume}{306}},
  \bibinfo{pages}{36} (\bibinfo{year}{1983}).

\bibitem{Kralj2012}
\bibinfo{author}{Kralj, J.~M.}, \bibinfo{author}{Douglass, A.~D.},
  \bibinfo{author}{Hochbaum, D.~R.}, \bibinfo{author}{Maclaurin, D.} \&
  \bibinfo{author}{Cohen, A.~E.}
\newblock \bibinfo{journal}{\bibinfo{title}{Optical recording of action
  potentials in mammalian neurons using a microbial rhodopsin}}.
\newblock {\emph{\JournalTitle{Nat Meth}}} \textbf{\bibinfo{volume}{9}},
  \bibinfo{pages}{90--95} (\bibinfo{year}{2012}).

\bibitem{Hochbaum2014}
\bibinfo{author}{Hochbaum, D.~R.} \emph{et~al.}
\newblock \bibinfo{journal}{\bibinfo{title}{All-optical electrophysiology in
  mammalian neurons using engineered microbial rhodopsins}}.
\newblock {\emph{\JournalTitle{Nat Meth}}} \textbf{\bibinfo{volume}{11}},
  \bibinfo{pages}{825--833} (\bibinfo{year}{2014}).

\bibitem{Wilt2013}
\bibinfo{author}{Wilt, B.~A.}, \bibinfo{author}{Fitzgerald, J.~E.} \&
  \bibinfo{author}{Schnitzer, M.~J.}
\newblock \bibinfo{journal}{\bibinfo{title}{Photon shot noise limits on optical
  detection of neuronal spikes and estimation of spike timing}}.
\newblock {\emph{\JournalTitle{Biophysical Journal}}}
  \textbf{\bibinfo{volume}{104}}, \bibinfo{pages}{51--62}
  (\bibinfo{year}{2013}).

\bibitem{Scanziani2009}
\bibinfo{author}{Scanziani, M.} \& \bibinfo{author}{Hausser, M.}
\newblock \bibinfo{journal}{\bibinfo{title}{Electrophysiology in the age of
  light}}.
\newblock {\emph{\JournalTitle{Nature}}} \textbf{\bibinfo{volume}{461}},
  \bibinfo{pages}{930--939} (\bibinfo{year}{2009}).

\bibitem{Keren2008}
\bibinfo{author}{Keren, K.} \emph{et~al.}
\newblock \bibinfo{journal}{\bibinfo{title}{Mechanism of shape determination in
  motile cells}}.
\newblock {\emph{\JournalTitle{Nature}}} \textbf{\bibinfo{volume}{453}},
  \bibinfo{pages}{475} (\bibinfo{year}{2008}).

\bibitem{Gauthier2012}
\bibinfo{author}{Gauthier, N.~C.}, \bibinfo{author}{Masters, T.~A.} \&
  \bibinfo{author}{Sheetz, M.~P.}
\newblock \bibinfo{journal}{\bibinfo{title}{Mechanical feedback between
  membrane tension and dynamics}}.
\newblock {\emph{\JournalTitle{Trends in Cell Biology}}}
  \textbf{\bibinfo{volume}{22}}, \bibinfo{pages}{527--535}
  (\bibinfo{year}{2012}).

\bibitem{Pierre2015}
\bibinfo{author}{Pierre, S.} \& \bibinfo{author}{Julie, P.}
\newblock \bibinfo{journal}{\bibinfo{title}{Membrane tension and cytoskeleton
  organization in cell motility}}.
\newblock {\emph{\JournalTitle{Journal of Physics: Condensed Matter}}}
  \textbf{\bibinfo{volume}{27}}, \bibinfo{pages}{273103}
  (\bibinfo{year}{2015}).

\bibitem{Zhang2001}
\bibinfo{author}{Zhang, P.-C.}, \bibinfo{author}{Keleshian, A.~M.} \&
  \bibinfo{author}{Sachs, F.}
\newblock \bibinfo{journal}{\bibinfo{title}{Voltage-induced membrane
  movement}}.
\newblock {\emph{\JournalTitle{Nature}}} \textbf{\bibinfo{volume}{413}},
  \bibinfo{pages}{428--432} (\bibinfo{year}{2001}).

\bibitem{Holthuis2014}
\bibinfo{author}{Holthuis, J. C.~M.} \& \bibinfo{author}{Menon, A.~K.}
\newblock \bibinfo{journal}{\bibinfo{title}{Lipid landscapes and pipelines in
  membrane homeostasis}}.
\newblock {\emph{\JournalTitle{Nature}}} \textbf{\bibinfo{volume}{510}},
  \bibinfo{pages}{48--57} (\bibinfo{year}{2014}).

\bibitem{Savtchenko2017}
\bibinfo{author}{Savtchenko, L.~P.}, \bibinfo{author}{Poo, M.~M.} \&
  \bibinfo{author}{Rusakov, D.~A.}
\newblock \bibinfo{journal}{\bibinfo{title}{Electrodiffusion phenomena in
  neuroscience: a neglected companion}}.
\newblock {\emph{\JournalTitle{Nat Rev Neurosci}}}
  \textbf{\bibinfo{volume}{18}}, \bibinfo{pages}{598--612}
  (\bibinfo{year}{2017}).

\bibitem{Petrov2002}
\bibinfo{author}{Petrov, A.~G.} \& \bibinfo{author}{Sachs, F.}
\newblock \bibinfo{journal}{\bibinfo{title}{Flexoelectricity and elasticity of
  asymmetric biomembranes}}.
\newblock {\emph{\JournalTitle{Physical Review E}}}
  \textbf{\bibinfo{volume}{65}}, \bibinfo{pages}{021905}
  (\bibinfo{year}{2002}).

\bibitem{Hill1977}
\bibinfo{author}{Hill, B.}, \bibinfo{author}{Schubert, E.},
  \bibinfo{author}{Nokes, M.} \& \bibinfo{author}{Michelson, R.}
\newblock \bibinfo{journal}{\bibinfo{title}{Laser interferometer measurement of
  changes in crayfish axon diameter concurrent with action potential}}.
\newblock {\emph{\JournalTitle{Science}}} \textbf{\bibinfo{volume}{196}},
  \bibinfo{pages}{426--428} (\bibinfo{year}{1977}).

\bibitem{Fang-Yen2004}
\bibinfo{author}{Fang-Yen, C.}, \bibinfo{author}{Chu, M.~C.},
  \bibinfo{author}{Seung, H.~S.}, \bibinfo{author}{Dasari, R.~R.} \&
  \bibinfo{author}{Feld, M.~S.}
\newblock \bibinfo{journal}{\bibinfo{title}{Noncontact measurement of nerve
  displacement during action potential with a dual-beam low-coherence
  interferometer}}.
\newblock {\emph{\JournalTitle{Optics Letters}}} \textbf{\bibinfo{volume}{29}},
  \bibinfo{pages}{2028--2030} (\bibinfo{year}{2004}).

\bibitem{Akkin2009}
\bibinfo{author}{Akkin, T.}, \bibinfo{author}{Landowne, D.} \&
  \bibinfo{author}{Sivaprakasam, A.}
\newblock \bibinfo{journal}{\bibinfo{title}{Optical coherence tomography phase
  measurement of transient changes in squid giant axons during activity}}.
\newblock {\emph{\JournalTitle{Journal of Membrane Biology}}}
  \textbf{\bibinfo{volume}{231}}, \bibinfo{pages}{35--46}
  (\bibinfo{year}{2009}).

\bibitem{LaPorta2012}
\bibinfo{author}{Laporta, A.} \& \bibinfo{author}{Kleinfeld, D.}
\newblock \bibinfo{journal}{\bibinfo{title}{Interferometric detection of action
  potentials}}.
\newblock {\emph{\JournalTitle{Cold Spring Harbor Protocols}}}
  \textbf{\bibinfo{volume}{2012}}, \bibinfo{pages}{pdb.ip068148}
  (\bibinfo{year}{2012}).

\bibitem{Iwasa1980}
\bibinfo{author}{Iwasa, K.}, \bibinfo{author}{Tasaki, I.} \&
  \bibinfo{author}{Gibbons, R.}
\newblock \bibinfo{journal}{\bibinfo{title}{Swelling of nerve fibers associated
  with action potentials}}.
\newblock {\emph{\JournalTitle{Science}}} \textbf{\bibinfo{volume}{210}},
  \bibinfo{pages}{338--339} (\bibinfo{year}{1980}).

\bibitem{Kim2007}
\bibinfo{author}{Kim, G.~H.}, \bibinfo{author}{Kosterin, P.},
  \bibinfo{author}{Obaid, A.~L.} \& \bibinfo{author}{Salzberg, B.~M.}
\newblock \bibinfo{journal}{\bibinfo{title}{A mechanical spike accompanies the
  action potential in mammalian nerve terminals}}.
\newblock {\emph{\JournalTitle{Biophysical Journal}}}
  \textbf{\bibinfo{volume}{92}}, \bibinfo{pages}{3122--3129}
  (\bibinfo{year}{2007}).

\bibitem{Nguyen2012}
\bibinfo{author}{Nguyen, T.~D.} \emph{et~al.}
\newblock \bibinfo{journal}{\bibinfo{title}{Piezoelectric nanoribbons for
  monitoring cellular deformations}}.
\newblock {\emph{\JournalTitle{Nat Nano}}} \textbf{\bibinfo{volume}{7}},
  \bibinfo{pages}{587--593} (\bibinfo{year}{2012}).

\bibitem{Popescu2006}
\bibinfo{author}{Popescu, G.}, \bibinfo{author}{Ikeda, T.},
  \bibinfo{author}{Dasari, R.~R.} \& \bibinfo{author}{Feld, M.~S.}
\newblock \bibinfo{journal}{\bibinfo{title}{Diffraction phase microscopy for
  quantifying cell structure and dynamics}}.
\newblock {\emph{\JournalTitle{Optics Letters}}} \textbf{\bibinfo{volume}{31}},
  \bibinfo{pages}{775--777} (\bibinfo{year}{2006}).

\bibitem{Bhaduri2014}
\bibinfo{author}{Bhaduri, B.} \emph{et~al.}
\newblock \bibinfo{journal}{\bibinfo{title}{Diffraction phase microscopy:
  principles and applications in materials and life sciences}}.
\newblock {\emph{\JournalTitle{Advances in Optics and Photonics}}}
  \textbf{\bibinfo{volume}{6}}, \bibinfo{pages}{57--119}
  (\bibinfo{year}{2014}).

\bibitem{Goetz2018}
\bibinfo{author}{Goetz, G.} \emph{et~al.}
\newblock \bibinfo{journal}{\bibinfo{title}{Interferometric mapping of material
  properties using thermal perturbation}}.
\newblock {\emph{\JournalTitle{Proceedings of the National Academy of
  Sciences}}} \textbf{\bibinfo{volume}{115}}, \bibinfo{pages}{E2499--E2508}
  (\bibinfo{year}{2018}).

\bibitem{Oh2012}
\bibinfo{author}{Oh, S.} \emph{et~al.}
\newblock \bibinfo{journal}{\bibinfo{title}{Label-free imaging of membrane
  potential using membrane electromotility}}.
\newblock {\emph{\JournalTitle{Biophysical Journal}}}
  \textbf{\bibinfo{volume}{103}}, \bibinfo{pages}{11--18}
  (\bibinfo{year}{2012}).

\bibitem{Batabyal2017}
\bibinfo{author}{Batabyal, S.} \emph{et~al.}
\newblock \bibinfo{journal}{\bibinfo{title}{Label-free optical detection of
  action potential in mammalian neurons}}.
\newblock {\emph{\JournalTitle{Biomedical Optics Express}}}
  \textbf{\bibinfo{volume}{8}}, \bibinfo{pages}{3700--3713}
  (\bibinfo{year}{2017}).

\bibitem{Schurmann2016}
\bibinfo{author}{Schürmann, M.}, \bibinfo{author}{Scholze, J.},
  \bibinfo{author}{Müller, P.}, \bibinfo{author}{Guck, J.} \&
  \bibinfo{author}{Chan, C.~J.}
\newblock \bibinfo{journal}{\bibinfo{title}{Cell nuclei have lower refractive
  index and mass density than cytoplasm}}.
\newblock {\emph{\JournalTitle{Journal of Biophotonics}}}
  \textbf{\bibinfo{volume}{9}}, \bibinfo{pages}{1068--1076}
  (\bibinfo{year}{2016}).

\bibitem{Steelman2017}
\bibinfo{author}{Steelman, Z.~A.}, \bibinfo{author}{Eldridge, W.~J.},
  \bibinfo{author}{Weintraub, J.~B.} \& \bibinfo{author}{Wax, A.}
\newblock \bibinfo{journal}{\bibinfo{title}{Is the nuclear refractive index
  lower than cytoplasm? validation of phase measurements and implications for
  light scattering technologies}}.
\newblock {\emph{\JournalTitle{Journal of Biophotonics}}}
  \textbf{\bibinfo{volume}{10}}, \bibinfo{pages}{1714--1722}
  (\bibinfo{year}{2017}).

\bibitem{Locquin2013}
\bibinfo{author}{Locquin, M.} \& \bibinfo{author}{Langeron, M.}
\newblock \emph{\bibinfo{title}{Handbook of Microscopy}}
  (\bibinfo{publisher}{Elsevier Science}, \bibinfo{year}{2013}).

\bibitem{Yang2018}
\bibinfo{author}{Yang, Y.} \emph{et~al.}
\newblock \bibinfo{journal}{\bibinfo{title}{Imaging action potential in single
  mammalian neurons by tracking the accompanying sub-nanometer mechanical
  motion}}.
\newblock {\emph{\JournalTitle{ACS Nano}}} \textbf{\bibinfo{volume}{12}}
  (\bibinfo{year}{2018}).

\bibitem{Pham2013}
\bibinfo{author}{Pham, H.~V.}, \bibinfo{author}{Edwards, C.},
  \bibinfo{author}{Goddard, L.~L.} \& \bibinfo{author}{Popescu, G.}
\newblock \bibinfo{journal}{\bibinfo{title}{Fast phase reconstruction in white
  light diffraction phase microscopy}}.
\newblock {\emph{\JournalTitle{Applied Optics}}} \textbf{\bibinfo{volume}{52}},
  \bibinfo{pages}{A97--A101} (\bibinfo{year}{2013}).

\bibitem{Litke2004}
\bibinfo{author}{Litke, A.~M.} \emph{et~al.}
\newblock \bibinfo{journal}{\bibinfo{title}{What does the eye tell the brain?:
  Development of a system for the large-scale recording of retinal output
  activity}}.
\newblock {\emph{\JournalTitle{Ieee Transactions on Nuclear Science}}}
  \textbf{\bibinfo{volume}{51}}, \bibinfo{pages}{1434--1440}
  (\bibinfo{year}{2004}).

\bibitem{Hottowy2012}
\bibinfo{author}{Hottowy, P.} \emph{et~al.}
\newblock \bibinfo{journal}{\bibinfo{title}{Properties and application of a
  multichannel integrated circuit for low-artifact, patterned electrical
  stimulation of neural tissue}}.
\newblock {\emph{\JournalTitle{Journal of Neural Engineering}}}
  \textbf{\bibinfo{volume}{9}}, \bibinfo{pages}{17} (\bibinfo{year}{2012}).

\bibitem{Park2014}
\bibinfo{author}{Park, J.} \emph{et~al.}
\newblock \bibinfo{journal}{\bibinfo{title}{Screening fluorescent voltage
  indicators with spontaneously spiking hek cells}}.
\newblock {\emph{\JournalTitle{PLOS ONE}}} \textbf{\bibinfo{volume}{8}},
  \bibinfo{pages}{e85221} (\bibinfo{year}{2014}).

\bibitem{McNamara2016}
\bibinfo{author}{Mcnamara, H.~M.}, \bibinfo{author}{Zhang, H.},
  \bibinfo{author}{Werley, C.~A.} \& \bibinfo{author}{Cohen, A.~E.}
\newblock \bibinfo{journal}{\bibinfo{title}{Optically controlled oscillators in
  an engineered bioelectric tissue}}.
\newblock {\emph{\JournalTitle{Physical Review X}}}
  \textbf{\bibinfo{volume}{6}}, \bibinfo{pages}{031001} (\bibinfo{year}{2016}).

\bibitem{Hosseini2016}
\bibinfo{author}{Hosseini, P.} \emph{et~al.}
\newblock \bibinfo{journal}{\bibinfo{title}{Pushing phase and amplitude
  sensitivity limits in interferometric microscopy}}.
\newblock {\emph{\JournalTitle{Optics Letters}}} \textbf{\bibinfo{volume}{41}},
  \bibinfo{pages}{1656--1659} (\bibinfo{year}{2016}).

\bibitem{Richards2005}
\bibinfo{author}{Richards, M.~A.}
\newblock \emph{\bibinfo{title}{{Fundamentals of Radar Signal Processing}}}
  (\bibinfo{publisher}{McGraw-Hill}, \bibinfo{year}{2005}).

\bibitem{Berlind2008}
\bibinfo{author}{Berlind, T.}, \bibinfo{author}{Pribil, G.~K.},
  \bibinfo{author}{Thompson, D.}, \bibinfo{author}{Woollam, J.~A.} \&
  \bibinfo{author}{Arwin, H.}
\newblock \bibinfo{journal}{\bibinfo{title}{Effects of ion concentration on
  refractive indices of fluids measured by the minimum deviation technique}}.
\newblock {\emph{\JournalTitle{physica status solidi (c)}}}
  \textbf{\bibinfo{volume}{5}}, \bibinfo{pages}{1249--1252}
  (\bibinfo{year}{2008}).

\bibitem{Catterall2017}
\bibinfo{author}{Catterall, W.~A.}, \bibinfo{author}{Wisedchaisri, G.} \&
  \bibinfo{author}{Zheng, N.}
\newblock \bibinfo{journal}{\bibinfo{title}{The chemical basis for electrical
  signaling}}.
\newblock {\emph{\JournalTitle{Nature Chemical Biology}}}
  \textbf{\bibinfo{volume}{13}}, \bibinfo{pages}{455} (\bibinfo{year}{2017}).

\bibitem{Rappaz2005}
\bibinfo{author}{Rappaz, B.} \emph{et~al.}
\newblock \bibinfo{journal}{\bibinfo{title}{Measurement of the integral
  refractive index and dynamic cell morphometry of living cells with digital
  holographic microscopy}}.
\newblock {\emph{\JournalTitle{Optics Express}}} \textbf{\bibinfo{volume}{13}},
  \bibinfo{pages}{9361--9373} (\bibinfo{year}{2005}).

\bibitem{Boss2013}
\bibinfo{author}{Boss, D.} \emph{et~al.}
\newblock \bibinfo{journal}{\bibinfo{title}{Measurement of absolute cell
  volume, osmotic membrane water permeability, and refractive index of
  transmembrane water and solute flux by digital holographic microscopy}}.
\newblock {\emph{\JournalTitle{Journal of Biomedical Optics}}}
  \textbf{\bibinfo{volume}{18}} (\bibinfo{year}{2013}).

\bibitem{Juffmann2016}
\bibinfo{author}{Juffmann, T.}, \bibinfo{author}{Klopfer, B.~B.},
  \bibinfo{author}{Frankort, T. L.~I.}, \bibinfo{author}{Haslinger, P.} \&
  \bibinfo{author}{Kasevich, M.~A.}
\newblock \bibinfo{journal}{\bibinfo{title}{Multi-pass microscopy}}.
\newblock {\emph{\JournalTitle{Nature Communications}}}
  \textbf{\bibinfo{volume}{7}}, \bibinfo{pages}{5} (\bibinfo{year}{2016}).

\end{thebibliography}

\clearpage



\end{document}